\documentclass{article}

\usepackage{arxiv}

\usepackage[utf8]{inputenc} 
\usepackage[T1]{fontenc}    
\usepackage{hyperref}       
\usepackage{url}            
\usepackage{booktabs}       
\usepackage{amsfonts}       
\usepackage{nicefrac}       
\usepackage{microtype}      
\usepackage{lipsum}
\usepackage{graphicx}
\usepackage{graphicx}
\graphicspath{ {./images/} }

\title{A Theoretical Framework of the Processes of Change in Psychotherapy Delivered by Artificial Agents}

\author{
 Arthur Bran Herbener \\
  Department of Psychology and Behavioral Sciences\\
  Aarhus University\\
  \texttt{abh@psy.au.dk} \\
   \And
 Malene Flensborg Damholdt \\
  Department of Psychology and Behavioral Sciences\\
  Aarhus University\\
  \texttt{malenefd@psy.au.dk} \\
}

\begin{document}
\maketitle
\begin{abstract}
The question of whether artificial agents (e.g., chatbots and social robots) can replace human therapists has received notable attention following the recent launch of large language models. However, little is known about the processes of change in psychotherapy delivered by artificial agents. To facilitate hypothesis development and stimulate scientific debate, the present article offers the first theoretical framework of the processes of change in psychotherapy delivered by artificial agents. The theoretical framework rests upon a conceptual analysis of what active ingredients may be inherently linked to the presence of human therapists. We propose that human therapists’ ontological status as human beings and sociocultural status as socially sanctioned healthcare professionals play crucial roles in promoting treatment outcomes. In the absence of the ontological and sociocultural status of human therapists, we propose what we coin the genuineness gap and credibility gap can emerge and undermine key processes of change in psychotherapy. Based on these propositions, we propose avenues for scientific investigations and practical applications aimed at leveraging the strengths of artificial agents and human therapists respectively.  We also highlight the intricate agentic nature of artificial agents and discuss how this complicates endeavors to establish universally applicable propositions regarding the processes of change in these interventions.
\end{abstract}


\section{Introduction}
What, if anything, is lost if we replace human therapists with artificial agents? The urgent nature of this question has only intensified in recent years. The hitherto unprecedented prevalence of mental health problems and extensive waiting lists for access to care has created a demand for new approaches to mental healthcare (1). In response to this pressing need, it has been suggested that artificial agents such as chatbots, digital avatars, and social robots can revolutionize the delivery of mental healthcare (2). Over the last decade, numerous scientific studies have shown promising effectiveness of evidence-based psychological treatments such as cognitive behavioral therapy (CBT) delivered by artificial agents in improving depression, anxiety, and well-being (3).  Additionally, some artificial agents are now capable of engaging in interactions nearly indistinguishable from human conversations, as seen in applications like ChatGPT, Replika, and Google Gemini (4). Research suggests that not even trained psychotherapists can reliably differentiate between therapeutic conversations led by human therapists and such artificial agents  (5). Thus, the traditional notion of “talk therapy” as an interpersonal endeavor may soon be subject to change.

However, as technology progresses swiftly and promising findings continue to emerge, essential questions remain unanswered, such as how it affects the therapeutic process if human therapists are replaced with artificial entities (6). While artificial agents are socially responsive technologies that can give the impression of mental states, intentionality, and empathy, critics stress that they inherently lack these qualities due to their ontological nature as probability-driven algorithmic devices (7, 8). This paradox – where artificial agents convincingly display human social behaviors yet remain fundamentally algorithmic – raises profound epistemic and ethical concerns for psychotherapy. In particular, the clinical consequences of this ambiguous nature remains uncertain, as does the appropriate terminology for psychological interventions led by artificial agents. That is, whether artificial agents can be meaningfully understood within the existing theoretical frameworks of psychotherapy, or if a new explanatory framework is needed (6).

The present article presents a theoretical framework of the change processes in psychotherapy delivered by artificial agents. The theoretical framework rests upon a conceptual analysis of what psychological processes may be inherently linked to the presence of the human therapist, and what this means for our understanding of psychotherapy delivered by artificial agents. Particularly, we underline the role of human therapists’ ontological status as human beings and sociocultural status as legitimate healthcare providers in promoting treatment outcomes. We argue how this status has implications not only for the understanding of how and to whom artificial agents may be effective, but also how the advantages of both artificial agents and human therapists respectively can be leveraged in praxis.

\subsection{Processes of Change in Psychotherapy}
A brief reflection on theory and evidence on the change processes in psychotherapy is necessary to clarify key premises of the proposed theoretical framework. There is a long-standing tradition for inquiring about what makes psychotherapy effective, a research field known as process-oriented research (9, 10). Process-oriented research is particularly interested in studying the active ingredients and mechanisms of action in psychotherapy. The concept of active ingredients refers to the components of a treatment that are believed to serve a remedying function (e.g., cognitive restructuring in CBT) (11). Active ingredients operate at a higher level of analysis than observable treatment characteristics, such as therapeutic techniques and conversational style (12). Active ingredients stimulate mechanisms of action, understood as the psychological processes that lead to symptomatic improvements (11, 12, 13), such as a decrease in dysfunctional cognitive beliefs.

Another differentiation is specific vs. common factors. These concepts represent different understandings of what makes psychotherapy effective. Specific factors refer to active ingredients that are unique to a particular treatment due to the distinct procedures, techniques, and tasks that differentiate it from other treatments (11), e.g., cognitive restructuring in CBT. Common factors are active ingredients that are not unique to a specific treatment due to commonalities in the structures, procedures, and forms of social interaction across therapies (10, 14), e.g., the therapeutic relationship and positive outcome expectations. Thus, a core difference lies in whether trajectories to change are believed to be unique or shared across treatments.

The present paper refrains from taking a stance on whether and which specific or common factors contribute most to treatment outcomes, given the inconclusive evidence. Research on specific factors often have null findings (10, 15, 16), and different treatments tend to demonstrate comparable outcomes across despite utilizing distinct treatment strategies (17, 18). However, methodological challenges (e.g., establishing temporal precedence and ensuring sufficient statistical power) complicate interpretations of these null findings (10, 12). In contrast, the therapeutic relationship (19) and positive outcome expectations (20) have been found to be the strongest predictors of treatment outcomes (10). Yet, there is limited time-lagged and experimental research on these common factors, challenging firm interpretations of causality. In broad terms, neither specific nor common factors perspectives have received sufficient convincing support  to be firmly accepted as explanatory frameworks for psychotherapy. Conversely, as neither approach has been sufficiently contested either, a cautious approach is warranted when constructing an analytical framework for psychotherapy delivered by artificial agents.

Nonetheless, common factors approaches in particular invite criticism towards the idea of psychotherapy delivered by artificial agents. For example, the contextual model – which arguably represents the most comprehensive and well-established modern common factor theory – proposes that psychotherapy works through three causal pathways: 1) a real relationship, which taps into basic human needs for belonginess and attachment; 2) expectations of improvement, wherein clients believe that the treatment will be helpful; 3) specific factors. Though, the model contends that therapeutic techniques are beneficial not because they target the disorder's etiological cause, but because therapeutic practices such as fostering more positive thinking styles are inherently healthy (10).

The contextual model’s emphasis on the therapeutic relationship and outcome expectations has implications for analyzing the change processes in psychotherapy delivered by artificial agents. It implies that the effectiveness of these interventions will rely on the degree to which humans can form real relationships to inanimate entities in a fashion that resembles the socio-emotional functions of interpersonal ones. It also implies that the beliefs clients’ hold about the credibility of artificial agents as effective therapists may influence treatment outcomes. In what follows, we will argue that such conditions are not easily manipulated by technological development, for example by designing artificial agents to comply with treatment protocols and adopting conversational styles akin to human therapists. As we will show, therapists’ ontological status as human beings and sociocultural status as healthcare professionals may have widespread implications for the therapeutic process, including the therapeutic relationship, the acquisition of social learning experiences, client outcome expectations, and client adherence.

\section{The Role of Ontological and Sociocultural Status}
\label{sec:headings}
The current pace of technological advancement suggests that the ability to emulate the verbal, and perhaps non-verbal, behaviors of human therapists are technological possibilities rather than metaphysical impossibilities. Nonetheless, the present article asserts that the disparity between the observable behaviors and the ontological and sociocultural status of artificial agents has the potential to undermine key change processes commonly believed to play significant roles to treatment outcomes. In this section, we argue that this disparity can result in what we coin the genuineness gap and credibility gap.

\subsection{The Genuineness Gap}
In social interactions, people can assume that others hold the internal capacities to think and feel. While making accurate inferences of others’ internal states can be challenging, there is generally no need to question the existence of their mental capacities based on folk theories of their ontological nature as human beings. This folk theory, however, does not necessarily apply to artificial agents. Although artificial agents can display signs of internal states such as consciousness and empathy, they lack the underlying mental structures enabling these internal states due to their ontological status as inanimate devices. The awareness of this ontological nature may have consequences for relationship development. For example, Seitz (21) found that  a chatbot’s empathetic actions tended to decrease user perceptions of authenticity, or self-congruence, in the chatbot. The effectiveness of social support from chatbots also appears to depend on implicit mind attributions to the chatbot. When humans do not implicitly attribute consciousness to a chatbot, they evaluate the helpfulness and effectiveness of its social support less favorably (22).

We conceptualize the gap between an artificial agent’s empathetic and friendly behaviors and awareness of its ontological nature as the genuineness gap. Drawing on theories of social cognition and the therapeutic relationship, we argue that the genuineness gap can undermine the socio-emotional functions of empathetic, validating actions from artificial agents. We assert that the genuineness gap may ultimately attenuate the development of a beneficial therapeutic relationship. (see Figure 1). Before we further elaborate on the genuineness gap, we briefly contemplate why interpersonal encounters in psychotherapy are important to clarify the relevance of the genuineness gap in a therapeutic context.

\begin{figure}
    \centering
    \includegraphics[width=0.75\linewidth]{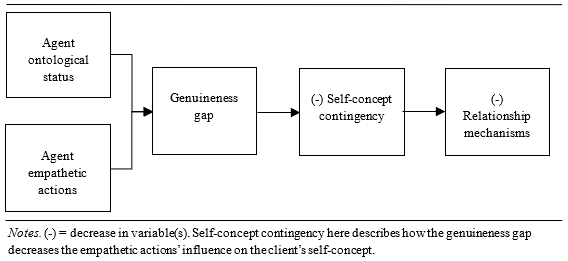}
    \caption{Causal Model of the Genuineness Gap}
    \label{fig1:placeholder}
\end{figure}

\subsubsection{Functions of the Therapeutic Relationship}
A classic and controversial question with relevance in general and to psychotherapy specifically is whether humans can form social relationships to artificial agents in a manner that resembles interpersonal ones. Answering this question is crucial, as therapist-client relationships, as well as interpersonal relationships in general, may have several socio-emotional benefits. First, according to the contextual modal, the client-therapist relationship may help satisfy basic human needs for social connectedness, a process thought to rely largely on therapists’ display of empathy (10). It is acknowledged that humans thrive when their social needs are fulfilled, while they may experience emotional distress when these needs are not met (23). As with social relationships in general, the therapist-client relationship holds the potential for fostering a sense of relatedness, which may be particularly beneficial for socially disconnected clients (10).

Second, a strong therapist-client relationship may facilitate corrective emotional experiences. This concept denotes a type of social learning experience occurring in interpersonal interactions in general and in psychotherapy specifically (24). It is characterized by the process wherein negative beliefs related to interpersonal relationships and interactions are challenged when engaging with someone who’s behavioral reactions misalign with one’s expectations  (25). For instance, a client who anticipates a negative reaction, such as condemnation, when disclosing something personal, may instead be met with empathy and acceptance, thereby challenging the belief that underlies the expectation. Corrective emotional experiences also explain why unconditional positive regard in Carl Roger’s client-centered approach might be helpful, given that the therapist’s accepting and non-judgmental attitude may foster a sense of self-worth (26). However, the perception that the therapist’s positive regard, empathy, and warmth reflect genuine sentiments might be a critical antecedent of these mechanisms, as we will show below.

\subsubsection{Antecedents of the Genuineness Gap}
Humans naturally infer causes of events and behaviors (27). Even when these inferences are erroneous, they foster a sense of predictability and control, as individuals become confident in their ability to anticipate future events and behaviors and navigate unfamiliar situations effectively. In interpersonal interactions, interpreting others’ internal states provides insights that help understand and predict their behaviors. Indeed, throughout the evolutionary history of humans, close relationships have formed between individuals, each with their own internal states, experiences, and narratives. Intersubjectivity, as understood by Stern (28), helps each participant in the dyadic encounter comprehend and share those experiences. By creating an understanding of the other’s intentions, attitudes, and feelings, each participant in the dyadic interactions gets to know and connect with each other and become assured that the other can be relied upon during times of distress (28, 29, 30). This approach aligns with contemporary perspectives provided by predictive coding theories of social cognition. In broad terms, predictive coding theory suggests that information conveyed through verbal and non-verbal communication is continuously integrated into cognitive representations of others’ mental states and sentiments (31). These representations allow individuals to infer others’ feelings, goals, and intentions to enable accurate predictions about their subsequent actions and facilitating assessments of interpersonal relationship quality.The idea that humans infer information about others’ internal states and use this information to evaluate their interpersonal relationships resonates with Gelso’s (2014) concept of the real relationship (10). Originally conceptualized as a feature of the therapeutic relationship and later serving as inspiration for the contextual model, the concept of the real relationship was developed to capture the kind of interpersonal relationship that goes beyond the professional-collaborative one between therapist and client. According to Gelso (2014), the real relationship is characterized by genuineness and realism, both marked by the degree to which they exist and by their valence. While genuineness signifies the perception that each person acts in a way that aligns with their authentic self (i.e., self-congruence), realism denotes the befitting perception of each other, i.e., that each part perceives the other in a way that fits who they are. Hence, the real relationship necessitates the belief that, for instance, the therapist’s positive regard is an authentic reflection of an accepting and non-judgmental stance based on a befitting understanding of one, rather than merely reflecting a façade (Gelso, 2014).

Gelso’s (2014) concept of the real relationship assumes a dualism between expressed behaviors and internal states, and this dualism has implications for the viability of artificial agents as relational partners. According to this understanding, in order for a relationship to be perceived as real, each participant in the dyadic interaction must believe there is a “something” within the counterpart that is capable of holding genuine sentiments. This idea aligns with the notion of intersubjectivity, as the ability is assumed to allow humans to make evaluations of what sentiments and intentions are responsible for others’ behaviors to foster an increased understanding of them (28). With artificial agents, however, there are no internal states to infer, considering their ontological nature as machines. The client is likely to be aware that the expressed empathy does not reflect a genuinely positive stance but is the product of statistical pattern-matching techniques and software developers’ design intentions. Therefore, the conceptual criteria for a real relationship, as conceptualized by Gelso (32), cannot be satisfied. As we shall see, being aware of an artificial agents’ ontological nature means that clients might not experience the same socio-emotional benefits as from relationships with human therapists.

\subsubsection{Implications of the Genuineness Gap}

The possible self-validating function of interpersonal relationships might help explain why the perception of genuineness is important. The notion that interpersonal relationships serve egocentric functions is not novel (33, 34). Initially coined by Sullivan (35), and more recently discussed by Wallace and Tice (36), the concept of reflected appraisal describes how the self-concept, understood as the overall comprehension of oneself, is shaped by interpretations of what others think of us. This process involves internalizing others' responses as reflections of one's value, competence, and identity, either reinforcing, maintaining, or challenging the self-concept. In close interpersonal relationships, reflected appraisals can foster closeness, intimacy, and security when feeling genuinely valued by the other (33). From an evolutionary perspective, reflected appraisal can be understood as a mechanism for survival and social cohesion. In this view, it enables individuals to gauge their social acceptability and adapt their behaviors to maintain belonging and support within their social group (37).

As in interpersonal interactions in general, reflected appraisals may also occur during therapy. First, a strong therapeutic relationship could foster a positive self-concept, as it informs the client that he/she is worth liking. Second, corrective emotional experiences can be conceptualized as a result of reflected appraisals leading to a more positive self-concept. For example, the client may interpret the therapist’s empathetic response as a sign that the therapist genuinely cares about the client – an experience that perhaps contrasts expectations grounded in earlier relationship and thereby fosters a more positive self-concept (25). 

As such, considering that reflected appraisals partly depend on interpretations of others’ internal states, the perceived degree of genuineness in others may be essential to how reflected appraisals affect peoples’ self-concept. In interpersonal interactions, perceived insincerity may on some occasions cast doubt on others’ true sentiments. In human-AI interactions, awareness that the artificial agent lacks the ability to hold positive sentiments may diminish the socioemotional benefits of its friendly, empathetic behaviors. Indeed, in Gelso’s (2014) perspective, the artificial agent’s empathetic behaviors do not signify that the client is a valuable, likeable person if the client is aware that the artificial agent does not possess the mental capacities for holding genuinely positive sentiments based on a befitting understanding of who the client is. Simply put, the artificial agent’s friendliness does not reflect a volitional stance rooted in a fellow human’s impression of the client’s character, so the expressed friendliness may not convey that the client is worth liking.

As a result, the client's self-concept may become less contingent on the artificial agent's empathetic and validating actions, potentially hindering relationship development and corrective emotional experiences. However, as we will discuss later (section 3.1 and 5.3), the extent to which humans perceive artificial agents as possessing or lacking a mind is not straightforward. People's phenomenological experiences of artificial agents vary and seem to depend on multiple factors across different levels of analysis. Before we further elaborate on that reservation, we highlight another factor that may influence the processes of change in psychotherapy delivered by artificial agents.

\subsection{The Credibility Gap}
In this section, we address a subject that has received little, if any, attention in the scientific literature: The credibility of artificial agents in the role as mental health providers. This inquiry is borne out of the notion that clients do not approach therapy as ‘blank slates’. On the contrary, they bring preconceived notions about what to expect from the therapeutic encounter, partly based on assumptions about the therapist’s ability to provide the necessary support (38, 39).

Understanding clients’ perception of artificial agents’ ability and willingness to help them – also known as therapist credibility (20) – is important, as it may influence their treatment adherence and their prognostic beliefs about treatment outcomes, also referred to as outcome expectations (40). According to Strong (39), a necessity to successful therapy is the client’s receptiveness to influence by the therapy. This receptiveness is influenced by the degree to which clients can discredit the therapist as an invalid source of information. To Strong (1968), the perceived expertness, trustworthiness, and attractiveness (or similarity to the client) counteract the discrediting of the therapist by signaling the therapist’s credibility in their position to influence the client. If therapists manage to establish their credibility, their ability to influence clients to commit to the treatment, challenge their cognitive beliefs, and change their habits increases accordingly, Strong (1968) proposed. The influential work of Frank and Frank (38) extends Strong’s (1968) propositions by arguing that therapists’ credibility, or ethos, can also help remoralize clients. In this perspective, a common characteristic of clients’ seeking therapy is a demoralized state resulting from a prolonged phase with repeated attempts to get better that have failed. Partly due to encountering a therapist that manages to convince the client about the helpfulness of the treatment, Frank and Frank (1991) suggest that a process of remoralization occurs, conceptualized as the mobilization of hope and self-efficacy.

What stands out from these perspectives is the idea that clients’ hold preconceptions about who is able to provide the necessary help and what capacities effective mental health providers must possess. This proposition has critical implications for understanding the viability of artificial agents in mental healthcare, as a growing body of research indicates that humans perceive these technologies as inferior in assuming certain roles traditionally held by humans (5, 41). While several studies indicate that humans tend to self-disclose more freely to artificial agents than to humans (42, 43), research also suggests a preference for human responses to social dilemmas, even when unknowingly rating AI responses as better (44); a preference to self-disclose personal issues to humans compared to AI specifically when seeking socioemotional support (45); and worse evaluations of empathy and helpfulness when believing that recommendations for relationship issues are written by a chatbot than a human therapist (46).

We propose that such observations reflect a credibility gap that can emerge from introducing artificial agents in the role of a mental health providers. We conceptualize this gap as the disparity between the observable performance of artificial agents in performing the role as mental healthcare providers and their potential lack of credibility in assuming this role. We argue this gap arises from cognitive prototypes shaped by sociocultural discourses about which personal and professional qualities are necessary to effectively deliver therapy (see Figure 2). 

\begin{figure}
    \centering
    \includegraphics[width=0.75\linewidth]{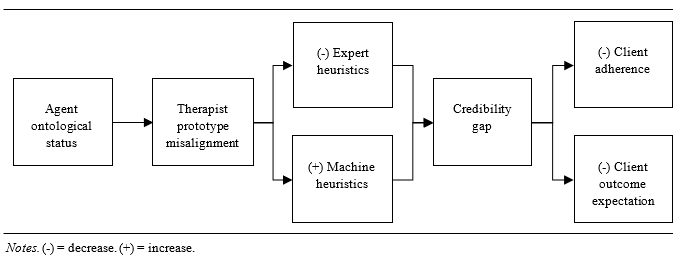}
    \caption{Causal Model of the Credibility Gap}
    \label{fig2:placeholder}
\end{figure}

\subsubsection{Functions of Therapists’ Sociocultural Status}
When software programmers design artificial agents to provide mental healthcare interventions, a key focus is to ensure that the artificial agent complies with a therapeutic framework and adopt a stance and conversational style believed to foster a strong therapeutic alliance. However, factors largely outside the control of software programmers might affect clients’ experience and expectations of the interaction as well. As emphasized by Strong (1968), perceived credibility is not only a matter of which competences therapists demonstrate. On the contrary, the perception of credibility, marked by factors such as expertness and trustworthiness, shapes clients’ expectations and experiences of the treatment. Additionally, according to (38), the actions and symbolic cues that convey credibility vary across cultures, depending on which professions hold the status and reputation as socially sanctioned and legitimate healthcare professionals. Which professions are reputable in the role as mental healthcare providers should be understood as a reflection of the dominant explanatory framework for making sense of the world, culturally and sub-culturally, whether that framework is secular and scientific, religious, or pseudoscientific.

Indeed, modern Western societies largely endorse practitioners of evidence-based practices, such clinical psychologists. Their expertise is underpinned by formal education, licensing, and adherence to professional and ethical standards (38). These qualities are signaled by professional credentials such as titles and diplomas, attire, a professional jargon, and physical settings such as private clinics and hospitals. In contrast, some traditional cultural contexts place greater trust in traditional healers, such as shamans, voodoo practitioners, or priests, who perform sacred rituals believed to address spiritual, psychological and somatic issues (47). According to Frank and Frank (1991), what unites physicians, clinical psychologists, and traditional healers across these cultural settings, despite their differing methods, is their status as socioculturally sanctioned and legitimate healing persons. In other words, their recognition as authorities and experts, and the institutionalization of their practice within a given culture. 

\subsubsection{Antecedents of the Credibility Gap}
Frank and Frank’s (1991) perspective implies that humans hold cognitive representations about what constitutes a typical therapist, and that these cognitive representations are shaped by cultural factors, including discursive practices. Such cognitive representations can be conceptualized as prototypes, a foundational principle in cognitive psychology that signifies a member of a conceptual category that embodies the common features shared by members of that category (48). Prototypes can influence categorization through representative heuristics that describes a tendency to assess categorical membership based on the perceived degree of similarity to a prototype (49). From this viewpoint, perceived credibility can be seen as how closely a therapist aligns with a client’s cognitive prototype of a credible therapist, also known as an instance of prototypicality. When a therapist exhibits sufficient traits that match this prototype, the client is more likely to perceive the therapist as credible.

At least two mechanisms might result from this prototype misalignment. Less favorable evaluations of artificial agents in roles traditionally held by humans may in some instances reflect less activation of expert heuristics. Expert heuristics are mental shortcuts believed to evoke credibility evaluations based on symbolic cues of expertise. (50), also described as the "expertise-credibility equation" (51). Although cognitive heuristics are typically associated with lower-order, automatic, and intuitive decision-making, Sundar and Shyam (51) suggest that expert heuristics can also influence higher-order, systematic decision-making. For instance, individuals may consciously use expertise-related cues when determining the credibility of presented information. Consistent with the notion of expert heuristics, research suggests that professional credentials serve an important role in credibility evaluations. When health-related messages are provided by doctors rather than lay persons, the information is generally perceived as more credible (52, 53). Research also indicates that cues of medical expertise (including professional attire with white coat) are associated with the highest credibility evaluations compared to business or casual dress (54). Accordingly, meta-analytical evidence suggests that therapists’ reputational cues (e.g., diplomas), attire, and technical jargon are associated with higher credibility evaluations and attitude and behavior change (55).

However, encounters with artificial agents might also activate heuristics unique to their ontological nature as machines, carrying affordances potentially linked to credibility evaluations through mechanisms distinct from those involved in credibility evaluations of human therapists. Machine heuristics denote mental shortcut with which human attribute machinic properties to artificial agents and similar technologies and use this attribution as the basis for evaluating their performance (56). Machine heuristics are often referred to as an explanatory framework for why humans evaluate the performance of machines variably across different tasks. The concept rests on the assumption that artificial agents generally are assigned to a different ontological category than humans, one associated with attributes distinct from those of humans based on folk theories about machines (41). Tasks believed to require objectivity, lack of prejudice, or the processing of extensive data may be considered more suitable for machines than for humans (56, 57). Conversely, in this understanding, tasks believed to require emotional comprehension and support may be more favorably evaluated when performed by a human due to a belief that artificial agents are unable to comprehensively understand emotions (44, 45, 58). 

The activation of machine heuristics may conflict with cues signaling credibility in the artificial agent, such as its appeal to a therapeutic framework or usage of a jargon typical for human therapists. This potential conflict may arise from a prototype of therapists that encompasses certain species-related ontological characteristics considered essential to effectively deliver therapy (e.g., emotional and rational comprehension, life experiences), shaped by sociocultural notions about what constitutes credible therapists (38).

\subsubsection{Implications of the Credibility Gap}
Prototypes of which professional and personal qualities characterize effective therapists likely entail that artificial agents will be perceived as inferior compared to healthcare professionals. Although the actual performance of artificial agents may become indistinguishable from that of humans, the limited institutional and public recognition of artificial agents may ultimately foster a credibility gap. As a consequence, clients’ beliefs about the usefulness of the interventions, as well as their receptiveness to the agents’ recommendations, can be at stake (39), ultimately compromising client adherence and expectations of improvement (38).

These propositions rest upon the premise that credibility is a social construction, and that artificial agents do not embody key symbolic cues signaling credibility in the role of mental health therapist, at least as this role is currently conceptualized in Western countries. Based on Foucault’s (59) idea of power structures, construing therapist credibility as a social construct implies that recognition by authoritarian institutions may play a crucial role in shaping artificial agents’ perceived credibility. For example, the process of gaining official recognition and institutionalization has been ongoing for several years for various internet-delivered CBT (iCBT) program. This might explain the high patient preference for iCBT, which approaches the levels observed for face-to-face interventions (60). Whether artificial agents will receive the empirical support necessary for positive recognition and institutionalization, however, cannot be assumed in advance.

However, a key limitation of our propositions thus far is that artificial agents have been evaluated using same criteria as human therapists – without considering the potential benefits related to the distinct characteristics of artificial agents. Despite artificial agents’ potential misalignment with prototypes of socially sanctioned healthcare providers, these technologies might also hold key strengths through their objectivity, consistent performance free of emotional biases, and the capability to rapidly access and integrate extensive evidence-based knowledge. The consequences of machine heuristics are therefore not straightforward. For example, among clients for whom objectivity and impartiality are particularly important (e.g., marginalized or stigmatized populations), the machine heuristic could facilitate more favorable credibility evaluations under some circumstances.

\section{A Theoretical Framework of the Change Processes in Psychotherapy Delivered by Artificial Agents}
Based on the theoretical reflection thus far, the present framework asserts that the ontological status as human beings and sociocultural status as healthcare professionals are critical underpinnings for the effectiveness of psychotherapy. Without these qualities, there is a risk for genuineness and credibility gaps, which may contest the development of a strong, beneficial therapeutic relationship, positive outcome expectations, and client adherence. Thus, the framework posits that the replacement of human therapists with artificial entities could fundamentally alter the epistemic landscape of psychotherapy by modulating specific and common factors.  For example, if clients’ outcome expectations are poor due to a perceived lack of credibility in the therapist, clients’ receptiveness to all therapeutic techniques might be at stake. Figure 3 depicts how the genuineness and credibility gaps are positioned within a generic conceptualization of the change process in psychotherapy.

Descriptive treatment components represent the artificial agents’ actions, such as the usage of therapy-specific strategies and techniques, as well as generic therapeutic techniques, such as active listening, self-disclosure, Socratic questioning, summarization, and validation.

The component active ingredients encompasses both specific and common factors. The framework refrains from assuming a stance in the debate about which active ingredients are more or less important, or which ingredients even exist. Rather, the framework applies to all theoretical approaches to psychotherapy (i.e., specific therapeutic approaches and common factor theories) in which therapist genuineness and credibility can be assumed to facilitate treatment outcomes.

Mechanisms of actions represent the psychological change processes resulting from the active ingredients, which in turn gives rise to therapeutic change. For example, the therapist’s positive regard of the client can be conceptualized as an active ingredient that can result in corrective emotional experiences. In the case of artificial agents, however, the positive regard may not serve its function as an active ingredient and therefore not lead to corrective emotional experiences, as argued earlier.
The component client moderators represent variables that influence the magnitude or direction of the relationship between treatment and outcome in general. This includes, but is not limited to, demographic variables, the specific psychiatric disorder, symptom severity, etiology, life events, as well as readiness to change.

The final component is attenuating conditions. Thus far, we have focused on the consequences of clients’ awareness of artificial agents’ ontological nature. However, there are variables that seem to decrease the perceived salience of artificial agents’ ontological nature and thereby attenuate the genuineness and credibility gaps. A comprehensive review of potential attenuating conditions is beyond the scope of this article, but it is worth noting that a higher-order process critical to hypothesis-development grounded in this framework could be anthropomorphization.

\begin{figure}
    \centering
    \includegraphics[width=0.75\linewidth]{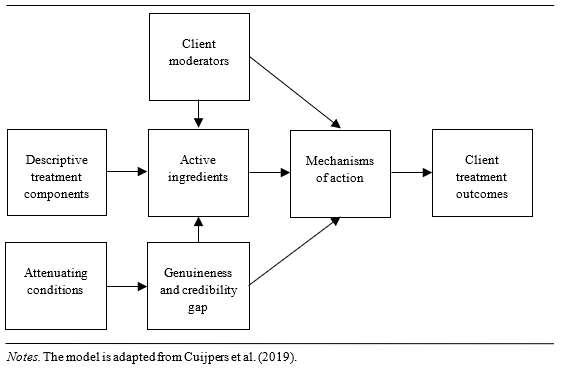}
    \caption{Integrative Model of the Processes of Change in Psychotherapy Delivered by Artificial Agents}
    \label{fig:placeholder}
\end{figure}

\subsection{The Attenuating Role of Anthropomorphization}
In the scientific literature, it has been argued that humans have an innate tendency for anthropomorphization. Although it has been critiqued for lack of conceptual clarity, anthropomorphization is commonly construed as the tendency to make attributions of mental traits, emotions or intentions to non-human animals and innate entities (61) This concept has been used to explain why some individuals engage in emotional relationships with chatbots and describe them as if they the chatbots genuinely understand and care about them (62, 63, 64). Thus, research on regular users of the social chatbot Replika has shown that anthropomorphism predicts an inclination to perceive social interactions with Replika as like interpersonal social interactions, which in turn predicted emotional attachment to Replika (65). There are also indications that anthropomorphism predicts attachment behaviours and psychological dependence to Replika (66). However, several factors at multiple levels of analysis appear to significantly contribute to the tendency to perceive artificial agents as possessing human-like mental capacities. At the individual level, potential factors include loneliness (67, 68), social anxiety (68), extraversion (69), and oxytocin levels (70). At the technological level, examples include human-like appearance (71), physical embodiment (72), conversational style (73) self-disclosure by artificial agents (78). At the cultural level, possible moderators include Shintoism and Animism  (74, 75). Thus, a coherent framework of psychotherapy delivered by artificial agents should embrace these three levels of variables that may influence anthropomorphization.

Cognitive processing theories suggest that anthropomorphism arises from an intuitive, fast, heuristic-based system (type 1 processing) that generates spontaneous social responses, even as a slower, more analytical system (type 2 processing) reminds users of the agent’s underlying mechanistic nature (76). This dual-process framework reveals a critical tension: while anthropomorphic cues can lower cognitive load and promote engagement, they may also foster over-attribution of human-like understanding, thereby obscuring the agent’s algorithmic basis. This underlines an essential distinction: The ontological properties of artificial agents as algorithmic devices vs. humans’ differential phenomenological experiences of them. Cultural factors further modulate these effects, as differences in societal narratives around psychotherapy and cultural influences on anthropomorphism could impact the perception of artificial agents as social entities (75)

Despite advances in the understanding of anthropomorphism, significant uncertainties persist. Notably, the interplay between machine heuristics and anthropomorphic processing remains poorly understood. The inter-relation between these perspectives is likely complex. In some designs, emphasizing the AI’s machine identity by highlighting data-driven insights might strengthen the machine heuristic and reduce emotional engagement. Conversely, emphasizing human-like qualities might trigger anthropomorphic trust but could reduce the sense of the AI’s computational objectivity. 

\section{Discussion and Future Directions}
Building on the theoretical framework presented above, it is important to explore the potential avenues for future research and practical applications that arise from our propositions. As we look ahead, it is clear that clinical trials incorporating assessments of mediators are essential to investigating and refining the proposed causal models. As of now, the propositions of the framework have only received indirect empirical support, primarily from analogous research, where external validity remains questionable.

Apart from the predictions that can be derived from this framework, it also opens up new directions for implementing artificial agents in mental healthcare. However, effective implementation may also require a deeper understanding of what works for whom and under which circumstances. We now turn to a discussion of scientific and practical implications of this framework. We also highlight some unresolved issues that should be taken into account during hypothesis-development grounded in this framework.

\subsection{What Works for Whom?}
The proposed theoretical framework gives rise to asking whether psychotherapy delivered by artificial agents may yield similar outcomes for all individuals or if some may benefit more than others. There are no easy answers to these questions. Specific therapeutic approaches often implicitly assume a one-size-fits-all model, presuming that all individuals within a given diagnostic category follow similar causal trajectories to therapeutic change (10). For example, traditional CBT assumes that depression is alleviated by identifying and challenging distorted thinking patterns through a structured, rational learning process (77). An underlying assumption of the one-size-fits-all model is that the presence of specific active ingredients is necessary to effectively treat certain client populations  (10). This assumption is based on the premise that individuals within diagnostic categories have similar treatment needs. Consequently, a one-size-fits-all approach implies that artificial agents in particular will be less efficacious in delivering treatments that rely heavily on active ingredients susceptible to genuineness and credibility gaps.

However, two key issues challenge this conclusion: a) research indicates that different therapeutic approaches often show similar effectiveness in treating specific disorders, a phenomenon known as the dodo bird verdict (10, 18); b) there is limited empirical support for the idea that each treatment operates through distinct causal mechanisms (10, 12, 78); Partly owing to such issues, several scientists endorse the common factors approach, often advocating for the essential role of the therapeutic relationship in general (79).

Still, although the therapeutic relationship seems to be the strongest predictor of treatment outcomes (10, 19), several issues prevent the conclusion that it is inherently and equally necessary for all clients. First, the existing evidence, given its correlational nature (19), does not establish causal relationships. Second, the observed correlations are based on group-level measurements and therefore provide limited grounds for interpreting inter-individual differences. Third, the mechanisms underlying the observed correlations remain unclear (12). Finally, it is uncertain whether the therapeutic relationship operates similarly across individuals. Indeed, the observed relationships can have different meanings. For some, it may signal a strong collaborative relationship. For others, the therapeutic relationship itself may play a direct causal role in driving treatment outcomes. Finally, for some, a strong therapeutic relationship may be an outcome of symptom reduction, reflecting reverse causality (10).

The emphasis on individual differences was also articulated by Kiesler (80) in what he coined the uniformity myth, encapsulating his criticism of the assumption that individuals within diagnostic categories follow similar trajectories to therapeutic change (80, 81). Skepticism towards the uniformity myth has led to research traditions focusing on what works for whom, including aptitude-treatment interaction research (82) and, more recently, personalized psychotherapy research. Within the latter tradition, Huibers et al. (81) proposed the personalized causal pathways hypothesis. This hypothesis states that treatment outcomes are determined by the match between the treatment and client characteristics. Although the hypothesis is somewhat vague, it prompts scientists to consider individual characteristics at a more fine-grained level of analysis than diagnostic categories, such as demography, symptom severity, etiology, cognitive characteristics, and life situation. This emphasis is also evident in the contextual model, which posits that clients with deficits related to interpersonal relationships, such as a history of chaotic relationships, particularly may benefit from the therapeutic relationship (10). In this perspective, a personalized approach implies that the genuineness gap may have different consequences across clients. Specifically, it suggests that artificial agents should not be implemented in individuals for whom the therapeutic relationship itself can be assumed to serve a key causal role in driving treatment outcomes. Similarly, the credibility gap could have varying consequences across clients, as research suggests that therapist credibility may be particularly important when pre-treatment expectations are low (83).

The notion that clients have different treatment needs has important implications for hypothesis-development grounded in our theoretical framework. Specifically, the potential for client-treatment interactions warrants careful consideration of which client groups and therapeutic frameworks are suitable for implementation in artificial agents. But it is also important to note that, although the personalized causal pathways hypothesis has received some support in certain contexts, there still lacks research that directly connects treatment components and outcomes across individuals (81). As such, expanding our understanding of client-treatment interactions is critical for hypothesis-development and implementation of artificial agents.

\subsection{Artificial Agents as Tools for Human Therapists?}
It must be stressed that artificial agents hold significant strengths in terms of scalability and accessibility. These strengths are underscored in comparison to the inherent limitations of the contemporary model of treatment in Western countries. That is, one-to-one sessions delivered by highly educated, well-paid healthcare professionals often located in urban areas (1). This treatment model is intrinsically linked to key barriers to care, such as waiting lists, economic costs, geographical distances, time, and stigma (1, 84, 85). Artificial agents can help address such issues. However, fully replacing mental health therapists come with potential losses in efficacy (perhaps more for some individuals than others), as anticipated by the proposed theoretical framework. So, how can we leverage the respective strengths of artificial agents and human therapists?

The answer perhaps lies in novel ways of allocating clients for treatment and tailoring interventions. As argued above, personalized psychotherapy is one potential avenue. The blended care model also represents a significant opportunity to leverage the ‘human factor’ and credibility associated with contemporary mental healthcare while capitalizing on the scalability and accessibility of artificial agents. Blended care, sometimes also called technology-supported care, is broadly characterized by mixing treatment components from different modalities, such as online exercises coupled with face-to-face sessions (86). An example of a contemporary, real-world application of blended care is internet-delivered interventions with therapist guidance. Internet-delivered interventions typically comprise structured online exercises adapted from evidence-based mental health treatments which clients (or patients) complete at home (87). The contact with therapists can be asynchronous, involving ongoing email contact, or synchronous, with face-to-face, online, or chat sessions. The idea behind therapist guidance is that the therapists’ role is to facilitate adherence and retention and provide feedback, while the online exercises deliver the active ingredients of the intervention (88).

Internet-delivered interventions can serve as inspiration for implementing artificial agents in a way that minimizes their potential disadvantages, including the genuineness and credibility gaps. However, the idea of implementing artificial agents in a blended care approach raises not only the question of what works for whom, but also in what blend? (86). In other words, what should the format, proportion, timing, and purpose be of therapist and artificial agent sessions, respectively? Our theoretical framework suggests benefits of incorporating therapist sessions in early treatment stages. By letting the therapist deliver the first treatment sessions, the therapist preserves the opportunity to build a strong therapeutic relationship that may directly impact treatment outcomes and contribute to the credibility of the treatment. Indeed, when the blended treatment is administered by a socially sanctioned healthcare professional, with whom the client ideally has a trusting relationship, the artificial agent gains a stamp of approval and may, consequently, capitalize on the credibility of the therapist. This, in turn, could enhance client retention and adherence to the technology-powered treatment component. On the other hand, the artificial agent’s sessions can play a consolidating function aimed at persistently engaging the client in utilizing the coping strategies taught by the therapist, accompanied by occasional follow-up therapist sessions.

Furthermore, a blended approach could be a low-risk strategy for gaining clinical experience and client feedback on artificial agents until a deeper understanding of their potential and limitations is in place. However, it is important to note that a blended approach places demands on therapists to learn, adapt, and embrace integrating technology into their practice Additionally, since the legal and moral responsibility for the artificial agent's performance also rests with the therapist (89), it is necessary to better understand how client interactions with artificial agents might have derivative implications for the client-therapist relationship. Such circumstances need to be considered when reflecting on the feasibility of a blended approach.

\subsection{Psychotherapy or something else?}
We must highlight the issues with applying the term ‘psychotherapy’ to refer to the scenario where artificial agents deliver evidence-based treatments in conjunction with therapeutic frameworks. By using this term, there is a risk of falling into the jingle fallacy where different phenomena are erroneously assigned to the same conceptual class due to sharing labels (90). In the present case, the jingly fallacy can yield the impression that psychological treatments delivered by humans and artificial agents are similar in terms of clinical change processes, which our framework contests. This is particularly an issue given our assertion that the therapeutic relationship – something that is often considered an integral component of psychotherapy (10, 12, 79)– may not evolve and operate in a similar fashion with artificial agents. Accordingly, a key conceptual feature across several definitions of psychotherapy is that it is performed by a therapist (91, 92, 93), and some scientists go as far as to suggest that psychotherapy is essentially a socially situated healing praxis (10, 38). From these perspectives, using the term psychotherapy presupposes an agentic status of artificial agents – understood as something that humans perceive, behave, and relate to as if it is human, as opposed to a utilitarian tool akin to a dishwasher or self-help book.

However, the agent vs. tool debate is an intricate matter. As proposed by Sedlakova and Trachsel (7), artificial agents have a hybrid nature, lying somewhere in-between the poles of agents and tools due to their ontological nature coupled with their human-like behavioral capacities. Furthermore, as discussed earlier, multiple variables appear to influence behaviors, perceptions, and relationships with artificial agents, such as visual features (e.g., humanoid appearance) and individual differences in loneliness and the tendency for anthropomorphism. Thus, the extent to which these technologies are attributed agency is neither fixed nor uniform across individuals and contexts. Even if one accepts the criterion that the presence of a human is necessary to label something as psychotherapy, what counts as human presence is not straightforward, phenomenologically speaking. For simplicity, we chose to use the label psychotherapy in this article. However, given our uncertainty as to what term best captures what is going on in interventions delivered by artificial agents, we welcome efforts to arrive at a more suitable term to help conceptually streamline this field.

\section{Concluding Remarks}
Exploring the implications of replacing human therapists with artificial entities has become crucial with the rise of large language models. To the best of our knowledge, the proposed theoretical framework is the first to offer tentative answers to how the introduction of artificial agents in the role of mental health therapists influence the change processes in psychotherapy. We assert that the ability to seamlessly emulate therapists’ verbal and non-verbal behaviors, despite being a matter of technological advancement in the foreseeable future, does not suffice to fully reenact the change processes of psychotherapy. We make this assertion by appealing to the possible therapeutic functions of human therapists’ ontological status as human beings and sociocultural status as healthcare professionals. In the absence of these basic characteristics, there emerges a risk of genuineness and credibility gaps, which may undermine the therapeutic value of the artificial agent’s actions.

A key contribution of the proposed theoretical framework is a conceptual foundation for hypothesis-development for guiding future research on the viability of artificial agents in psychotherapy. However, as a theory addressing a highly novel phenomenon, it is likely that our assertions will become subject to refinement as this research field progresses. This is partly due to the quite preliminary comprehension of human-AI interaction. Longitudinal experimental research studying relationship processes between humans and artificial agents is limited, as is research on the therapeutic relationship to artificial agents specifically (6). In particular, the understanding of anthropomorphism is limited. This is not only in terms of its core conceptual features but also regarding the underlying cognitive and perceptual processes, and how the interaction between anthropomorphism and awareness of an agent's ontological nature influences relationship development and socio-emotional outcomes. Therefore, while the proposed theoretical framework supports hypothesis-development, key issues remain unresolved in foundational research on human perceptions of artificial agents, especially concerning future embodied agents with humanoid morphology.

Long-standing issues in psychotherapy research also has implications for the validity of the proposed theoretical framework. Despite the nearly century-long search for active ingredients (9), the processes responsible for therapeutic change remain notoriously uncertain, particularly due to methodological challenges (10). As a consequence, our framework’s underlying mechanistic assumptions are still subject to debate. Expanding our general comprehension of psychotherapy is therefore key to establishing how, to whom, and under which circumstances artificial agents are effective in the context of therapy. Yet, it is important to note that artificial agents offer a promising venue for learning more about psychotherapy in general. Due to the resource demands of psychotherapy research, a crucial issue is to achieve sufficient statistical power to delineate subtle effects, an issue that is exacerbated by the potential of multitude of variables collectively responsible for change. Artificial agents offer a highly scalable platform enabling large-scale studies aimed at detecting even subtle effects. Furthermore, this technology may also help circumvent the ethical challenges that have prohibited experimental manipulations of relational variables in psychotherapy (10). Thus, research on artificial agents may generate insights that can be generalized to psychotherapy more broadly.

Hence, the present article represents one of the initial, exploratory steps in an emerging journey toward uncovering the potentials and limitations associated with employing artificial agents in mental healthcare. A comprehensive understanding of this topic will demand sustained research efforts across various fields, encompassing both foundational and applied research. Particularly, it also requires scientists to grapple directly with fundamental questions about human nature – specifically, understanding what it is about interacting with “others” rather than mere “things” that fosters meaningful emotional bonds. Adding complexity to the study of human-AI interaction is the accelerating pace of technological innovation, which continually reshapes the landscape. Consequently, a core responsibility of researchers in this research field is to raise and address basic questions regarding human nature that extend beyond any specific technological advancement, ensuring their relevance remains intact even as artificial agents continue to evolve. Such efforts are crucial to extend our understanding of artificial agents in psychotherapy specifically and human social arenas more broadly.

\bibliographystyle{unsrt}  


\end{document}